# Spatial-to-Temporal Orbital Angular Momentum Mapping in Twisted Light Fields

Vijay Kumar and Purnesh Singh Badavath

Department of Physics, National Institute of Technology Warangal, Telangana – 506004
`vijay@nitw.ac.in`

**Abstract.** Optical angular momentum (OAM) in light beams is manifested as the two-dimensional spatial distribution of its complex amplitude, necessitating a 2D detector for its measurement. Here we present a novel speckle-based machine learning approach for OAM recognition, which enables recognition using a 1D array detector or even a 0D single-pixel detector.

## 1 Introduction

The exponential surge in communication data bandwidth demand is driven by the proliferation of internet-connected devices, high-definition video streaming, cloud computing, and 5G technology. To support the requisite gigabits per second (Gbps) data rates, modern networks are exploring advanced modulation schemes, wavelength-division multiplexing (WDM), space division multiplexing (SDM), and machine learning-assisted network management. SDM, utilizing higher-order modes such as orbital angular momentum (OAM) beams, significantly enhances the information capacity of free-space links. This approach not only boosts bandwidth but also improves robustness against environmental disturbances, paving the way for reliable, high-capacity optical communication networks. Consequently, OAM beams offer a promising solution for meeting the escalating data transmission demands, ensuring efficient and high-speed communication on a global scale.

Numerous OAM demultiplexing techniques exist, and all exhibit significant sensitivity to alignment and noise [1]. Even AI-based OAM demultiplexing methods necessitate capturing the entire 2D mode for accurate identification [2]. Recently, machine learning-assisted speckle-based demultiplexing has demonstrated resistance to noise and misalignment by effectively recognizing OAM modes through a small segment of the speckle field [3-11]. The core concept of machine learning-assisted speckle-based OAM demultiplexing lies in identifying OAM modes using their corresponding speckle patterns rather than the modes themselves. These speckle patterns contain hidden information about the OAM modes, enabling efficient classification via a supervised deep-learning model. Existing AI-based OAM demultiplexing methods rely on the intensity of the modes, which fails to distinguish intensity-degenerate modes [2]. In contrast, speckle-based OAM demultiplexing leverages speckle patterns, allowing for the classification of intensity-degenerate OAM modes as well [6-9].

In Section 2, we will first introduce the classification of OAM modes using 2D speckle images. This will be followed by a discussion of OAM classification with a 1D spatial speckle array in Section 3.1. Finally, Section 3.2 will cover the methods for performing OAM classification using a single-pixel detector.

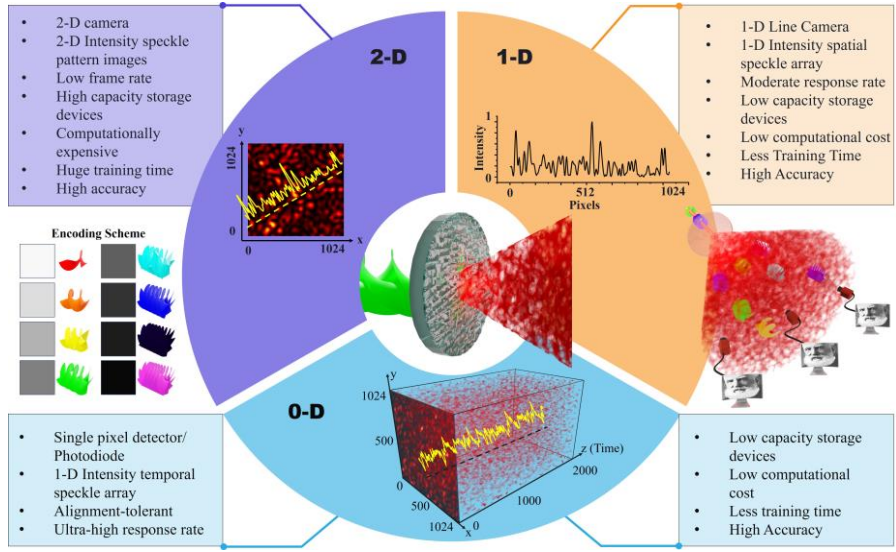

**Fig. 1.** Speckle-based OAM demultiplexing using 2D, 1D and 0D detectors.

## 2      2D Speckle-based OAM demultiplexing using a 2D detector

The 2D speckle-based OAM demultiplexing technique involves generating a speckle field by passing a propagated OAM beam through a diffuser, as illustrated in Fig. 1. The resulting 2D speckle pattern is captured using a 2D detector, such as a CMOS or CCD camera, positioned at the far-field. These captured speckle pattern images are then fed to an optimized Convolutional Neural Network (CNN), which is trained to recognize the incoming OAM beams based on their corresponding speckle patterns [3-11].

To classify intensity-degenerate OAM beams, astigmatism is introduced into the generated OAM speckle field using a cylindrical lens [6-9] and information transfer through astigmatic speckle-learned OAM shift-keying and OAM multiplexing was successfully demonstrated [8,9]. Experimental results indicated that astigmatic speckle pattern images captured using a 1f Fourier transforming system provided better classification accuracy than those captured using a 2f Fourier transforming system. [9].

Aiming to miniaturize OAM demultiplexing devices, the 2D speckle-based demultiplexing was demonstrated using randomly grown ZnO nanosheets on an aluminum substrate with nanoscale inhomogeneities. The generated speckle pattern due to the interaction of OAM beams with the ZnO nanosheets was extensively studied from near- to far-field using a 4f imaging system. The trained custom-designed CNN was able to classify OAM beams in both transverse and longitudinal directions, making it the most alignment-independent 2D speckle-based OAM demultiplexing technique [10].

To further reduce computational cost and enhance alignment independence, a 1D speckle-based OAM demultiplexing method is also proposed in the spatial and temporal domains.

## 3  1-D Speckle-based OAM demultiplexing

### 3.1  OAM demultiplexing using a 1-D array detector or line camera

The 1-D Speckle-based OAM demultiplexing in the spatial domain employs a 1D speckle array mapped along a random direction from the captured 2D speckle pattern image, as shown in Fig. 1. A 1D speckle array can be experimentally capture directly using a line camera. A thorough study has been done on the length of the array, and the number of observations per class affects the training time and accuracy. It has been reported that for an array of length 1×1920 pixels and 2000 observations per class, the custom-designed 1-D CNN has achieved 98% classification accuracy [11]. Thereby enhancing the demultiplexing speed and effectively reducing both the computational cost and training time. As an application for the proposed model, a non-line-of-sight communication channel has been established by employing the structured light shift keying encoding scheme. Further, the encoded information has been demultiplexed by 1-D speckle-based demultiplexing scheme [5, 11].

In this section, we have demonstrated that OAM demultiplexing can be achieved using a 1-D array detector or a line camera. In the next section, we will show that even a 0-D or single-pixel detector is sufficient for recognizing OAM beams.

### 3.2  OAM demultiplexing using 0-D or single-pixel detector

Addressing the challenge of recognizing OAM at higher rates, crucial for coping with the ever-increasing data transmission demands, remains a significant concern for researchers and industry professionals. Here we propose a novel method to enhance OAM recognition rates by employing a single-pixel detector. Our approach leverages one-dimensional speckle-based OAM demultiplexing in the temporal domain to achieve OAM detection at higher speeds. The one-dimensional temporal speckle data, extracted from the simulated and experimentally captured speckle pattern images in an uncorrelated manner, is subsequently fed into a Support Vector Machine (SVM) model for training. The trained SVM model successfully recognized the simulated and experimentally captured intensity degenerate OAM beams ($l=\pm 8$, $p=0$) with an accuracy exceeding 97.1% and 97.3% respectively [12].

## 4  Conclusion

In conclusion, the speckle-based machine learning approach for OAM recognition can be experimentally realized using a 2D or 1D or even a 0D detector, offering several advantages. It is alignment-free in both transverse [3-9] and longitudinal directions [10], does not require the entire modal field [3-6], and can classify intensity degenerate modes [6-9]. This method is effective from near-field to far-field and can be implemented using micro- to nano-structures [10]. Additionally, it is reference-free, single-shot, scalable to all families of structured light beams, and requires low computational cost [11]. Furthermore, it enables non-line-of-sight or omnidirectional structured light communication [5,7].

**Acknowledgement:** SERB India funding (SRG/2021/001375)

## References

1. Willner, A. E., Pang, K., Song, H., Zou, K. and Zhou, H.: Orbital angular momentum of light for communications. Appl Phys Rev 8, 041312-1 (2021).
2. Avramov-Zamurovic, S., Esposito, J. M. and Nelson, C.: Classifying beams carrying orbital angular momentum with machine learning: tutorial. J. Opt. Soc. Am. A 40(1), 64–77 (2023).
3. Raskatla, V., Badavath, P. S., Patil, S., Kumar, V., Singh, R. P.: Speckle-based Deep Learning Approach for OAM Modes Classification. J. Opt. Soc. Am. A, 39, 759 (2022).
4. Raskatla, V., Badavath, P. S., Kumar, V.: Convolutional Networks for speckle-based OAM modes classification. Optical Engineering 61, 036114 (2022).
5. Badavath, P. S., Raskatla, V., Chakravarthy, P., and Kumar, V.: Speckle-based Structured Light Shift-keying for Non-line-of-sight Optical Communication. Appl. Opt., 62, G53 (2023).
6. Raskatla, V., Badavath, P. S., Kumar V.: Speckle-learned Convolutional Neural Network for the recognition of intensity degenerate orbital angular momentum modes. Optical Engineering 62, 36104 (2023).
7. Raskatla, V., Badavath, P. S., Kumar, V.: Machine learning meets singular optics: speckle-based structured light demultiplexing. In: Sixteenth International Conference on Correlation Optics 2024, Proc. SPIE vol. 12938, pp. 129381H, SPIE (2024).
8. Das, T., Pandit, M. R., Raskatla, V., Badavath, P. S., Kumar, V.: Astigmatic speckle-learned OAM shift keying and OAM multiplexing. J Opt (2024). https://doi.org/10.1007/s12596-024-01899-7
9. Sharma, C., Badavath, P. S., and Kumar, V.: Experimental realization of optical communication link with intensity degenerate orbital angular momentum beams. J Opt (2024). https://doi.org/10.1007/s12596-024-01984-x
10. Sharma, C., Badavath, P. S., Potu, S., Rajaboina, R. K., and Kumar, V.: Machine Learning-assisted Orbital Angular Momentum Recognition using Nanostructures. J. Opt. Soc. Am. A, 41, 1420 (2024).
11. Badavath, P. S., Raskatla, V., and Kumar, V.: 1D Speckle-learned Structured Light Recognition. Opt. Lett. 49, 1045 (2024).
12. Badavath, P. S., and Kumar, V.: Single-pixel orbital angular momentum detection in temporal domain. J. Opt. 27, 01LT01 (2024).